\newcommand{\ergs}{erg\,s$^{-1}$}

\newcommand{\Msol}{$M_{\odot}$}
\newcommand{\Rsol}{$R_{\odot}$}

\newcommand{\Lbol}{\rm L$_{\rm bol}$}
\newcommand{\Lx}{$\rm L_{\rm X}$}

\newcommand{\sao}{\object{SAO\,49725}}
\newcommand{\hd}{\object{HD\,161103}}
\newcommand{\gcas}{\object{$\gamma$\,Cas}}
\newcommand{\rmag}{$r_{m}$}
\newcommand{\rmaglim}{$r^{lim}_{m}$}
\newcommand{\rc}{$r_{c}$}
\newcommand{\Macc}{$\dot{M}_{acc}$}
\newcommand{\Macclim}{$\dot{M}^{lim}_{acc}$}
\newcommand{\Ps}{$P_{s}$}
\newcommand{\Porb}{$P_{orb}$}

\documentclass{aa} \usepackage{natbib,graphicx}  \usepackage{natbib,graphicx} 

\begin{document}


\title{New \gcas-like objects: X-ray and optical observations of \sao\ and \hd\thanks{Based on observations obtained with XMM-{\it Newton}, an ESA science mission with instruments and contributions directly funded by ESA Member
States and NASA.
}}

\author{R. Lopes de Oliveira\inst{1,2} \and C. Motch\inst{1}  \and F. Haberl\inst{3} \and I. Negueruela\inst{4} \and E. Janot-Pacheco\inst{2}}

\offprints{R. Lopes de Oliveira,\\
\email{rlopes@newb6.u-strasbg.fr}}

\institute{Observatoire Astronomique, UMR 7550 CNRS, 11 rue de l'Universit\'e, F-67000 Strasbourg, France \and Instituto de Astronomia,
Geof\'{\i}sica e Ci\^encias Atmosf\'ericas, Universidade de S\~ao Paulo, R. do Mat\~ao 1226, 05508-090 S\~ao Paulo, Brazil \and
Max-Planck-Institut f\"ur extraterrestrische Physik, Giessenbachstra{\ss}e, 85748 Garching, Germany \and Departamento de F\'{\i}sica, Ingenier\'{\i}a de Sistemas y Teor\'{\i}a de la Se\~nal, Escuela Polit\'ecnica Superior, Universidad de Alicante, Ap. 99, 03080 Alicante, Spain }

\date{Received / Accepted}

\authorrunning{Lopes de Oliveira et al.}  \titlerunning{New \gcas-like objects.}

\abstract{A growing number of early Be stars exhibit X-ray luminosities intermediate between those typical of
early type stars and those emitted by most Be/X-ray binaries in quiescence. 
We report on XMM-{\it Newton} observations of two such Be stars, \sao\ and \hd\ which were originally discovered in a systematic cross-correlation between the ROSAT all-sky survey and SIMBAD. 
The new observations confirm the X-ray luminosity excess (\Lx\,$\sim$\,10$^{32-33}$\,erg\,s$^{-1}$) and the hardness of their X-ray spectra which are both unusual for normal early type stars. An iron K$\alpha$ complex is clearly detected in \hd\ in which the H-like, He-like and fluorescent components are resolved, while strong evidences also exist for the presence of similar features in \sao.
X-ray spectra can be equally well
fitted by a thermal plasma ({\it mekal}) with T\,$\sim$\,10$^{8}$\,K and solar abundances or by a {\it
power law}\,+\,{\it iron line} model with photon index $\sim$\,1.5--1.8, both with a soft thermal component with T\,$\sim$\,10$^{7}$\,K. 
The intensity of the fluorescence 6.4 keV line reflects the presence of
large amounts of cold material close to the X-ray sources and strongly argues against accretion onto a companion neutron star in a large orbit.
On the other hand, the probable thermal origin of the X-ray emission as supported by the ionised iron lines is in  disagreement with those observed in all known Be/X-ray binaries, in which a non-thermal component is always required.
Remarkably, the X-ray features are similar to those of white dwarves in
several cataclysmic variables.
There is no evidence for high frequency
pulsations in none of the two systems.  However, a large oscillation in the light curve of \hd\ with P\,$\sim$\,3200\,s is readily detected.  
The X-ray light curve of \sao\ exhibits clear variability by  $\sim$\,80\% on time scales as short as $\sim$\,1000\,s.  
New optical observations provide updated spectral types (B0.5\,III-Ve), and disclose a dense, large and apparently stable circumstellar disc for both stars. 
The nature of the excess X-ray emission is discussed in the light of the models proposed for \gcas,
magnetic disc-star interaction or accretion onto a compact companion object -- neutron star or white dwarf. 
These two new objects added to similar cases discovered in XMM-{\it Newton} surveys point at the emergence of a new class of \gcas\ analogs.

 }

\maketitle

\section{Introduction}\label{introduction}

\sao\ and \hd\ are two B0.5\,III-Ve stars identified as relatively strong X-ray emitters through cross-correlation
between the ROSAT all-sky survey and O-B star catalogs \citep{Motch97}. The short ROSAT observation did not allow detailed spectral studies but showed that the two sources had harder X-ray spectra than those of normal O-B stars, and luminosities ($\sim$\,10$^{32}$\,erg\,s$^{-1}$ in the 0.1--2.4\,keV energy range) midway between those radiated by normal stars and classical Be/X-ray transients in quiescence as well as persistent low-luminosity Be/X-ray binaries (X\,Per-like objects). Based on these peculiarities, \cite{Motch97} suggested that \sao\ and \hd\ were likely new Be/X-ray binaries and owing to the modest X-ray luminosity possible Be+white dwarf (Be+WD) systems.

Be/X-ray binaries constitute the most numerous class of high mass X-ray binaries ($\sim$2/3 of them) in which gravitational
capture onto a compact object of the Be star circumstellar material accounts for the X-ray emission \citep[see][for a recent compilation of Be/X-ray properties]{Raguzova05}.
Be+WD systems are produced in large numbers by models describing the evolution
of massive binaries, in which 20\% to 70\% of the Be stars formed as result of the binary evolution must have a white dwarf as companion \citep[see e.g. ][]{vandenHeuvel87,Waters89a,Pols91,VanBever97,Raguzova01}.
In spite of their high expected frequency, no such Be+WD binary has been yet firmly established. The only tenable Be+WD candidate known so far is \gcas\ whose X-ray and optical properties are similar to those of \sao\ and \hd.
In almost all well studied Be/X-ray binaries the X-ray spectra, the peak luminosities, pulsations, spin up rates and magnetic fields point altogether or separately at a neutron star (NS). 

In this paper we present the spectral and timing analysis of new X-ray observations of \sao\ and \hd\ obtained by the XMM-{\it Newton} satellite and optical spectroscopy carried out during several observational campaigns. The X-ray and optical properties of \sao\ and \hd\ are notably close to those of \gcas. We discuss the nature of the X-ray emission in both objects in the light of the single star and binary models proposed for \gcas.

\section{Observations
}

\subsection{Optical observations}

Optical spectroscopy
 of \sao\ and \hd\ has been obtained with a variety of 
 telescopes and instruments, listed in Table\,\ref{journalopticalspectra}. 
 
 \sao\ was observed at the Observatoire de Haute Provence (OHP, France)
 with the 1.93-m telescope and the {\it Carelec} spectrograph on 1995 September 28
 and on 2001 
 October 22 and 24.
 Other observations were taken at OHP with the 1.52-m telescope and the {\it Elodie} 
 spectrograph between 2001 July 31 and August 2. 
 Four observing runs were performed with the 1.52-m 
 G.~D.~Cassini telescope at the Loiano Observatory (Bologna, Italy) on 1999 August 18, 2000 July 25, 2004 July 17 and in 2005 July 6. On all occasions, the telescope was equipped with the Bologna Faint Object Spectrograph and
 Camera (BFOSC). Several grisms were used, in order to cover both the 
 blue and H$\alpha$ regions at moderate resolutions ($\sim$\,3--4\,\AA). 
 Finally, data were taken with the 2.5-m Isaac 
 Newton Telescope (INT) at La 
 Palma (Spain) and the Intermediate Dispersion Spectrograph (IDS) on 2003 
 July 1 and 2. Details of the configurations are 
 shown in Table~\ref{journalopticalspectra}. 
 
 \begin{table} \caption{Journal of optical spectroscopic observations.} 
 \label{journalopticalspectra} 
 \centering 
 \begin{tabular}{cccc} 
 \hline 
 \hline 
 Date & Telescope &  Wavelength & Dispersion \\ 
  & & range (\AA) & (\AA/pixel)\\ 
 \hline 
\multicolumn{4}{c}{{\bf SAO\,49725}} \\
 \hline 
 
 28/09/95	& OHP 193	& 3750--6750 			& 6.9	\\
 18/08/99	& Cassini 	& 4100--6700 			& 1.9	\\ 
 25/07/00	& Cassini 	& 3100--5300 			& 1.6	\\ 
 25/07/00	& Cassini 	& 6150--8150 			& 1.6	\\ 
 31/07/01	& OHP 152 	& 3960--4410 			& 0.22	\\ 
 01/08/01	& OHP 152 	& 4460--4910 			& 0.22	\\ 
 02/08/01	& OHP 152 	& 6380--6830 			& 0.22 	\\ 
 22/10/01	& OHP 193 	& 6250--7120 			& 0.45	\\ 
 24/10/01	& OHP 193 	& 3800--5500 			& 0.90	\\ 
 01/07/03	& INT 		& 5300--7000			& 0.65	\\ 
 02/07/03	& INT 		& 3800--5500			& 0.65	\\ 
 17/07/04	& Cassini	& 3900--10200$^{\mathrm{a}}$	& 0.8	\\
 06/07/05	& Cassini	& 3900--10200$^{\mathrm{a}}$	& 0.8	\\
 \hline 
\multicolumn{4}{c}{{\bf HD\,161103}} \\
 \hline 
 13/02/94			& ESO\,1.5\,m	& 4050--7200	& 1.9	\\
 01/08/98			& INT		& 3670--5060	& 0.5	\\
 02/08/98			& INT		& 5800--7100	& 0.5 	\\
 03/08/98			& INT		& 7700--8950	& 0.5 	\\
 01...05/03/03$^{\mathrm{b}}$	& ESO\,3.6\,m	& 3600--5050	& 1.9	\\
 07/06/03			& NTT		& 3980--4765	& 0.8	\\
 07/06/03			& NTT		& 6320--7830	& 0.4	\\
 \hline 

 \end{tabular} 
 \begin{list}{}{}
\item[$^{\mathrm{a}}$] covered in different orders;
\item[$^{\mathrm{b}}$] five observations (one per night).
\end{list}

 \end{table} 
 
 \hd\ was observed in low dispersion with a Boller \& Chivens spectrograph mounted at the 1.5-m ESO on 1994 February 13.  Medium resolution spectra were acquired with the INT\,+\,IDS on 1998 August 1-3 (data retrieved from the archive). Observations were carried out during five consecutive nights in March 2003 using the ESO 3.6-m telescope equipped with the EFOSC instrument, and with the ESO-NTT at La Silla and the ESO Multi-Mode Instrument (EMMI) on 2003 June 7. 

\subsection{X-ray observations}

XMM-{\it Newton} observed \sao\  on 2003 December 9 during 11.2\,ks (ObsID 0201200201) and \hd\ on 2004 January 26 during 17.6\,ks (ObsID 0201200101). In both cases, the EPIC cameras were operated in {\it large window} mode providing a time resolution of 48\,ms and 0.9\,s for pn and MOS1-2 cameras respectively. The medium filter was used to reject optical light from the relatively bright counterparts.  All data were processed with Science Analysis Software version 6.1.0 using the most recent calibration files. Spectral analysis was performed with the {\it Xspec}-v11.3 package.

Because of the relatively large count rate of the two sources, the inclusion of times with background flares has little impact on their spectra and light curves. At E\,$\ga$\,0.5\,keV spectra accumulated from the whole observation do not significantly differ from those accumulated from low background intervals only. In the following we will discard times of high background for spectral analysis (except for the Fe line complex study; Section \ref{Felinecomplexsection}) but will consider the whole observation for time variability studies, excluding the low energy bands. 

\begin{figure*} \centering 
\includegraphics[bb=1.7cm 3.5cm 10.6cm 27.5cm,clip=true,angle=-90,width=18cm]{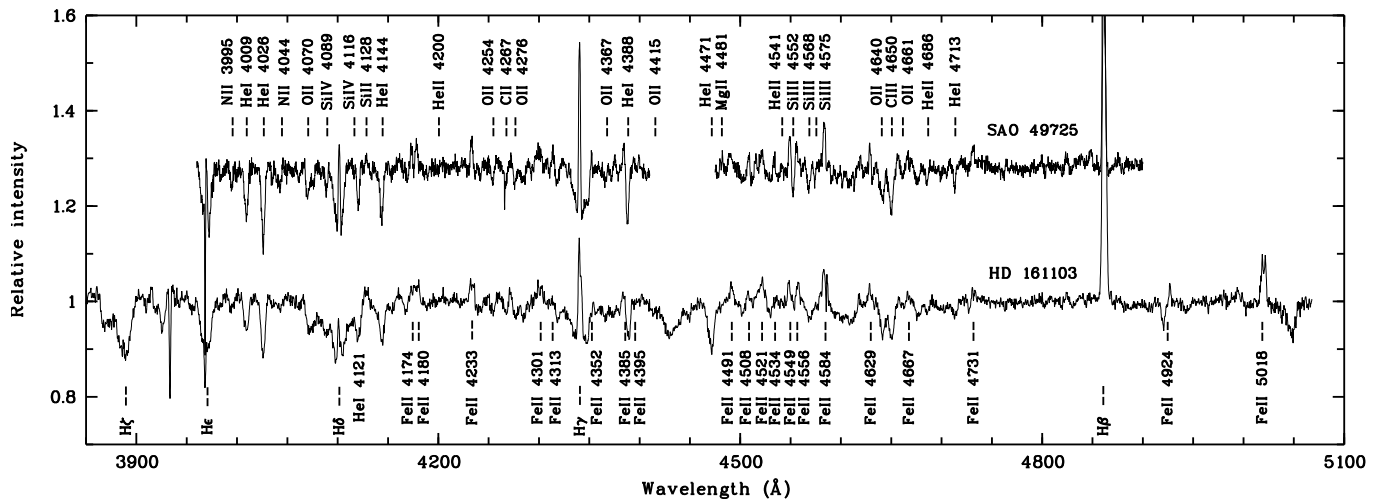} \caption{The optical spectra of \sao\  ($\lambda\lambda$3956--4412\AA~on 2001 July 31; $\lambda\lambda$4460--4915\AA~on 2001 August 1) and \hd\ (on 1998 August 1) in the classification region. See Table\,\ref{journalopticalspectra}.}
\label{opticalspectra} 
\end{figure*}

\begin{figure*}[] \centering 
\includegraphics[bb=0.7cm 0cm 18.5cm 28cm,clip=true,angle=-90,width=8.7cm]{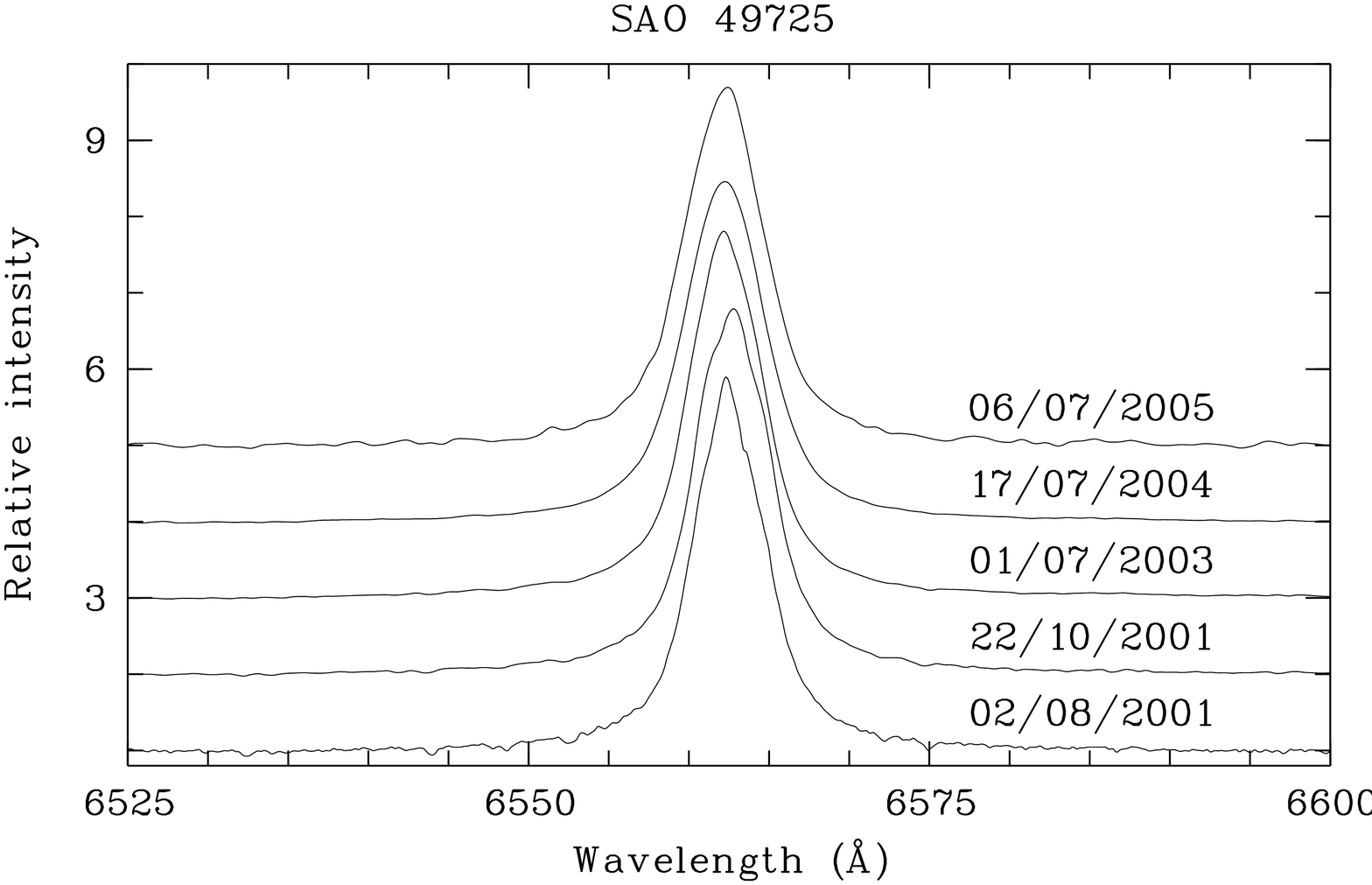} 
\includegraphics[bb=0.7cm 0cm 18.5cm 28cm,clip=true,angle=-90,width=8.7cm]{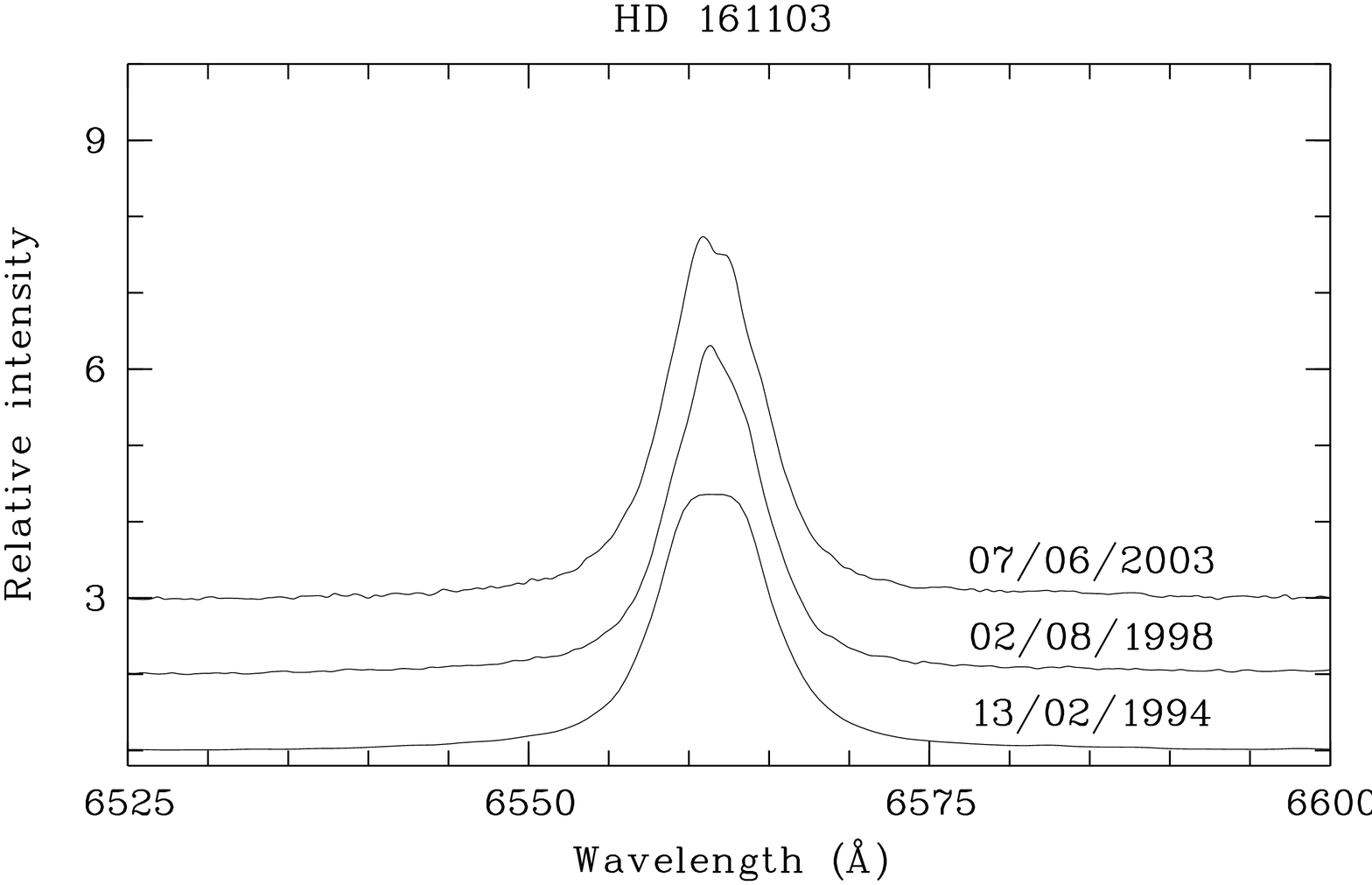} 
\caption{The H$\alpha$ profiles of \sao\ and \hd\ from observations with similar spectral resolutions. See Table\,\ref{journalopticalspectra} and Table\,\ref{opticalparameters}.}
\label{halphafig} 
\end{figure*}

The resulting exposure times in the flare free good time intervals were 9.2\,ks (MOS1 and MOS2) and 7.9\,ks (pn) for \sao, and 16.6\,ks (MOS1), 17.1\,ks (MOS2)  and 7.6\,ks (pn) for \hd.   In our analysis only events with pattern 0 to 4 and 0 to 12 were used for pn and MOS cameras respectively. Source counts were extracted in circular regions with radius of 35\arcsec~for \sao\ and 40\arcsec~for \hd\ centered on the X-ray emission. The backgrounds for each observation were measured in large boxes located close to the central source and on the same CCD. 

The position of the X-ray sources is fully consistent with those of the proposed optical counterparts. The angular distances of 0.7\arcsec~and 1.8\arcsec~between X-ray and optical positions of \sao\ and \hd\ respectively are well within the corresponding 90\% confidence error radii of 1.1\arcsec\ and 3.2\arcsec.

\section{Data analysis}

\subsection{Optical spectroscopy}
\label{opticalspectroscopysection}

\begin{figure*} \centering 
\includegraphics[bb=1.7cm 3.4cm 10.6cm 27.5cm,clip=true,angle=-90,width=18.4cm]{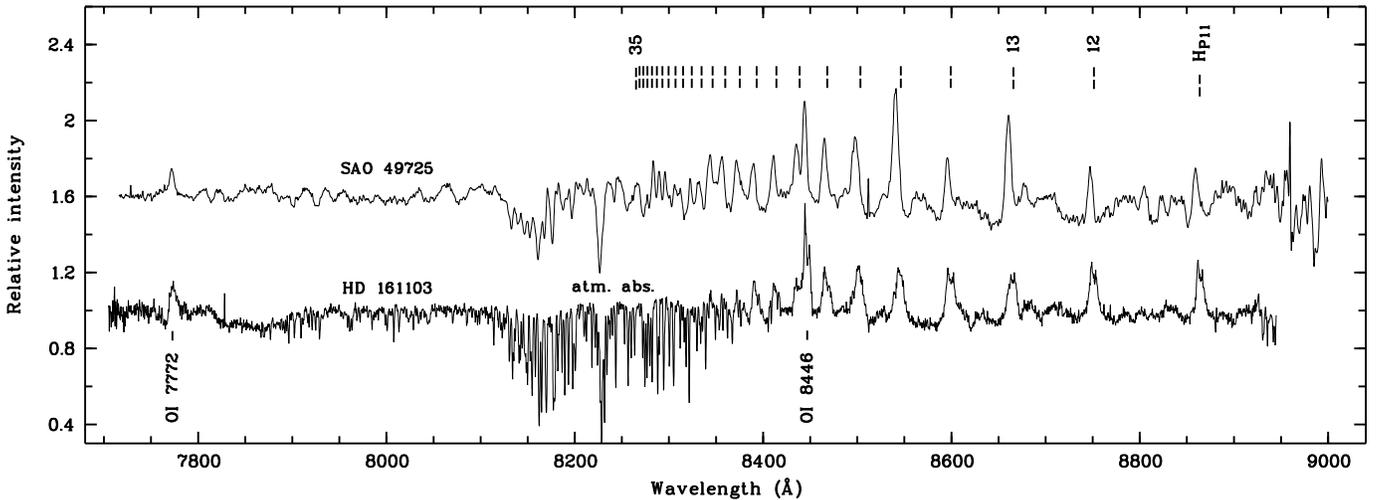} \caption{The optical spectra of \sao\ (on 2004 July 17) and \hd\ (on 1998 August 3) in the near infrared region. See Table\,\ref{journalopticalspectra}.}
\label{irsaohd} 
\end{figure*}

Figure\,\ref{opticalspectra} shows optical spectra of \sao\ and \hd\ in the $\lambda\lambda$4000--4900\,\AA\ region which is the best suited for spectral classification. The two objects have very similar spectra. A number of Fe\,{\small II} emission lines linked to the circumstellar environments are visible. No significant velocity change seems present during the 5 day long ESO 3.6-m observations of \hd\ in 2003. 
The radial velocities of the He\,{\small I}\,$\lambda$4026, He\,{\small I}\,$\lambda$4144, and He\,{\small I}\,$\lambda$4388 photospheric lines of \hd\ vary by less than 13\,km\,s$^{-1}$. 
For \sao\ this limit is 40\,km\,s$^{-1}$ for the He\,{\small I}\,$\lambda$4026 line, and 17\,km\,s$^{-1}$ for the He\,{\small I}\,$\lambda$4144 and He\,{\small I}\,$\lambda$4388 lines on a time scale of days or years. 
Rotational velocities were derived using the FWHM(He\,{\small I}\,$\lambda$4471) -- $vsini$ correlation of \citet{Slettebak75} \citep[see details in][]{Steele99}, resulting in a mean value of $vsini$\,=\,182\,$\pm$\,33\,km\,s$^{-1}$ and 234\,$\pm$\,34\,km\,s$^{-1}$ for \sao\ and \hd\ respectively.
\citet{Steele99} find a consistent value of  $vsini$\,=\,224\,$\pm$\,33\,km\,s$^{-1}$ using the He\,{\small I} lines in \hd.
\sao\ and \hd\ also show quite similar H$\alpha$ emission lines (see Fig.\,\ref{halphafig} and Table \ref{opticalparameters}), whose intensities and quasi-symmetrical profiles, both exhibiting little variability over long time intervals, suggest dense, large and probably stable circumstellar envelopes \citep{Slettebak92,Tycner05}. 
This stability contrasts 
with the rather strong variability displayed by most Be/X-ray binaries on 
timescales of years. Among the known exceptions are A\,1118--616 and Cep\,X--4, in which the accreting neutron stars may be in large and close to circular orbits, as suggested by \citet{Okazaki01}. 
The EW(H$\alpha$)--\Porb\ relation discovered by \citet{Reig97} in Be/X-ray binaries, if applicable to these systems, would suggest \Porb\,$\ga$\,100 days for \sao\ and \hd\ .

We also note that the strength of the N\,{\small II}\,$\lambda$3995, 4044 lines may indicate that \sao\ and \hd\ are moderately Nitrogen-rich, as are several X-ray binary primaries.

\begin{table}[top] \caption{Equivalent width of the H$\alpha$ emission line in \sao\ and \hd. The errors are estimated to be $\pm$\,1\,\AA.}              
\label{opticalparameters}       
\centering          
\begin{tabular}{c c c}  

\hline
\hline                    

Object & Date & EW(H$\alpha$) \\
& & (\AA) \\
\hline                    

\sao    	& 28/09/95   & --30		\\
		& 18/08/99   & --32	 	\\
		& 02/08/01   & --30	 	\\
    		& 22/10/01   & --32	 	\\
    		& 01/07/03   & --31	 	\\
		& 17/04/04   & --32		\\
		& 06/07/05   & --30		\\

\hline            	 

\hd\    	& 13/02/94   & --32		\\
		& 02/08/98   & --31		\\
		& 07/06/03   & --34		\\

\hline                    
\end{tabular} \end{table}

In both stars, the Si\,{\small III}\,$\lambda$4552 and Si\,{\small IV}\,$\lambda$4089 lines are consistent with a B0.5 type, which is also in agreement with the intensity of the He\,{\small II}\,$\lambda$4686 line seen in absorption. Later types are excluded by the weakness of the Si\,{\small II}\,$\lambda\lambda$4128-4130 and Mg\,{\small II}\,$\lambda$4481 lines. 
The O\,{\small I} triplet at $\lambda$7774 (72, 74 and 75) is in emission as is the Fe\,{\small II}\,$\lambda$7712 line, in agreement with an early type Be star classification.
The moderate emission of the metallic lines suggests, at least for \sao, a III-IV luminosity class, also supported by the strength of the O\,{\small II}\,$\lambda$4285 and Si\,{\small IV}\,$\lambda$4089 lines. 
Our classifications are consistent with the B0.5\,III-Ve spectral type derived by \citet{Motch97} for \sao\ and \citet{Steele99} for \hd.

\sao\ and \hd\ were also observed in the near infrared (Fig. \ref{irsaohd}). As expected for Be stars with intense H$\alpha$ line, their near-infrared spectra are dominated by the Paschen line series in emission, with positive detections up to H$_{P}$\,$\la$\,30. The resolution in both observations was not large enough to distinguish the Ca\,{\small II} triplet ($\lambda$8498/$\lambda$8542/$\lambda$8662) from the nearby Paschen lines. 
The O\,{\small I}$\lambda$8446 line is also seen in emission. Such feature is generally seen in Be stars to correlate well with H$\alpha$ line, as it is due to excitation by Ly$\beta$ photons \citep{Andrillat88}.
The Paschen and Balmer H$\alpha$ lines of \hd\ display evidences for a weak V/R asymmetry.

The distance to the two stars is somewhat uncertain owing to the lack of exact luminosity class and to the possible additional red contribution of the large circumstellar envelope to the intrinsic stellar colours. Assuming for \sao\ V\,=\,9.23, B--V\,=\,0.37 and for \hd\ V\,=\,8.7, B--V\,=\,0.43 \citep[][and references therein]{Motch97} and intrinsic colours and magnitudes in \citet{Wegner94} and \citet{Humphreys84} yields distances of 1.6 or 2.8\,kpc for \sao\ and 1.1 or 2.0\,kpc for \hd\ for luminosity classes V or III respectively. An additional conservative error of about 20\,\% on distances probably results from other uncertainties.

\subsection{X-ray spectroscopy}

\begin{figure*}
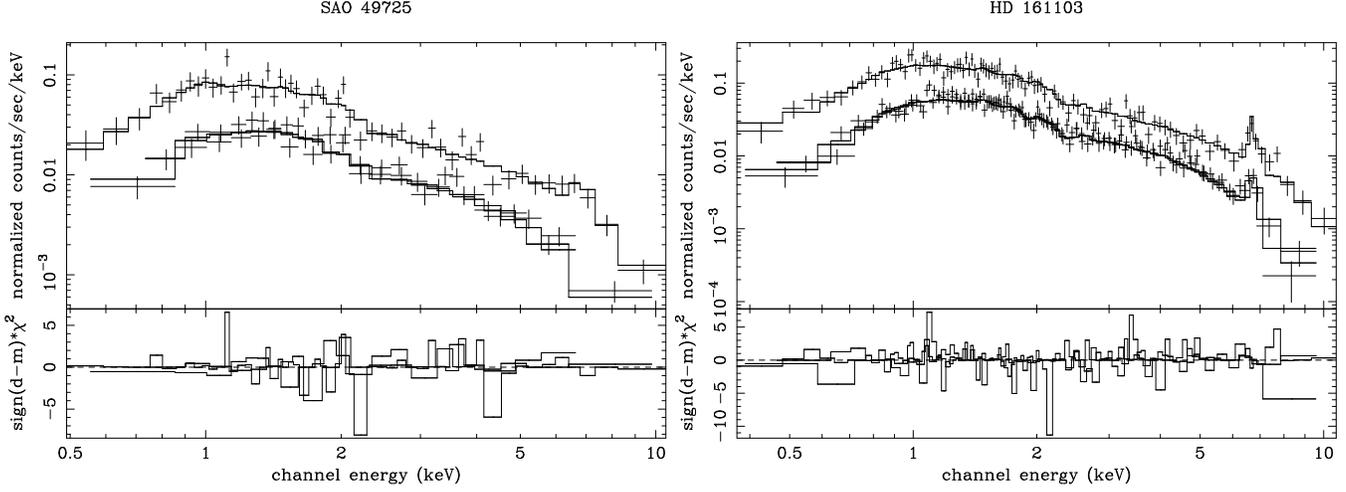
 \centering{ 
\includegraphics[angle=-90,width=8.8cm]{sao49725_wabs_mekalmekal_paper.ps}
\includegraphics[angle=-90,width=8.8cm]{hd161103_wabs_mekalmekal_paper.ps}
\caption{EPIC spectra of \sao\ and \hd.
The best fit absorbed {\it wabs}*({\it mekal}+{\it mekal}) model is shown for each individual EPIC camera. See Table \ref{fittingparameters}.} 
\label{xrayspectrum}  } \end{figure*}

\begin{table*} \caption{Best fit parameters for the X-ray spectra of \sao\ and \hd\ accumulated in time intervals free of background flares. Thermal models assume solar
abundances. Quoted errors are at the 90\% confidence level. Fluxes are given unabsorbed in the 0.2--12\,keV energy
band and averaged over the 3 EPIC cameras. We adopted the {\it wabs} model to account for the photo-electric absorption.}              
\label{fittingparameters}        \centering           \begin{tabular}{r c c c c c c c}            

\hline                    
\hline

& {\it Mekal} 	& \multicolumn{2}{c}{{\it 2-Mekal}} 	& {\it P.law\,+\,Gauss. line}	& {\it Mekal\,+\,P.law\,+\,Gauss. line}				\\
&		& {\it 1-absorption} & {\it 2-absorptions}	& 	& 									\\

\hline
\multicolumn{6}{l}{{\bf SAO\,49725}} \\
\hline

N$_\mathrm{H_{1}}$ (10$^{22}$\,cm$^{-2}$)		 & 0.30$^{+0.04}_{-0.04}$    &    0.35$^{+0.06}_{-0.06}$ &  	0.14$^{+0.11}_{-0.09}$  &  0.35$^{+0.06}_{-0.05}$      & 0.38$^{+0.09}_{-0.04}$ 	\\
$k$T$_{1}$ (keV)				 & 11.72$^{+5.66}_{-2.46}$   &    0.87$^{+0.50}_{-0.25}$ &  	0.93$^{+0.09}_{-0.11}$  & ...			       & 0.92$^{+0.47}_{-0.25}$ 	\\
EM$_{1}$$^{\mathrm{a}}$ (10$^{55}$\,cm$^{-3}$)	 & 2.6 (7.8)  		     &    0.02 (0.08)		 &  	0.02 (0.05)	        & ...			       & 0.02 (0.08)			\\
N$_\mathrm{H_{2}}$ (10$^{22}$\,cm$^{-2}$)		 & ...			     &    ...  			 &  	0.36$^{+0.05}_{-0.03}$  & ...			       & ...				\\
$k$T$_{2}$ (keV)				 & ...			     &    12.85$^{+7.73}_{-3.37}$&  	12.29$^{+1.52}_{-1.48}$ & ...			       & ...				\\
EM$_{2}$$^{\mathrm{a}}$ (10$^{55}$\,cm$^{-3}$)	 & ...			     &    0.8 (2.6)		 &      0.8 (2.6) 	        & ...			       & ...				\\
Photon index 					 & ...			     & ... 			 &   ...		        & 1.57$^{+0.12}_{-0 .06}$      & 1.53$^{+0.11}_{-0 .10}$	\\
Line (keV)						 & ...			     &  ... 			 &    ...		        & 6.61$^{+0.49}_{-0.51}$       & 6.61$^{+0.35}_{-0.36}$ 	\\
$\sigma_{\rm Line}$ (keV) & ...			     &  ... 			 &    ...		        & $\la$ 1.17		       & $\la$ 0.90			\\
f$_{\rm x}$ (erg\,cm$^{-2}$\,s$^{-1}$)		 & 1.25$\times$10$^{-12}$    &    1.30$\times$10$^{-12}$ &  	1.29$\times$10$^{-12}$  & 1.38$\times$10$^{-12}$       & 1.41$\times$10$^{-12}$ 	\\  
$\chi^{2}_{\nu}$/d.o.f.$^{\mathrm{b}}$		 & 1.11/94		     &    1.07/92		 &  	1.08/91 	        & 1.16/91		       & 1.13/89			\\

\hline
\multicolumn{6}{l}{{\bf HD\,161103}} \\		
\hline

N$_\mathrm{H_{1}}$ (10$^{22}$\,cm$^{-2}$)		 & 0.31$^{+0.02}_{-0.02}$  & 0.35$^{+0.02}_{-0.02}$   & 1.28$^{+0.24}_{-0.31}$   & 0.38$^{+0.03}_{-0.03}$	 & 0.40$^{+0.05}_{-0.04}$     \\
$k$T$_{1}$ (keV)				 & 7.44$^{+1.20}_{-0.63}$  & 0.76$^{+0.20}_{-0.17}$   & 0.20$^{+0.10}_{-0.06}$   & ...				 & 0.80$^{+0.85}_{-0.27}$     \\
EM$_{1}$$^{\mathrm{a}}$ (10$^{55}$\,cm$^{-3}$)	 & 3.2 (9.8)		   & 0.1 (0.3) 	      	      & 18.2 (55.8)	         & ...				 & 0.1 (0.3)		      \\
N$_\mathrm{H_{2}}$ (10$^{22}$\,cm$^{-2}$)		 & ...			   & ...  		      & 0.36$^{+0.06}_{-0.04}$   & ...				 & ...			      \\
$k$T$_{2}$ (keV)				 & ...			   & 8.01 $^{+1.05}_{-1.00}$  & 8.78 $^{+1.62}_{-1.22}$  & ...				 & ...			      \\
EM$_{2}$$^{\mathrm{a}}$ (10$^{55}$\,cm$^{-3}$)	 & ...			   & 1.5 (9.7) 		      & 3.1 (9.5)	         & ...				 & ...			      \\
Photon index 					 & ...			   & ... 		      & ...		         & 1.75$^{+0.07}_{-0 .06}$	 & 1.71$^{+0.07}_{-0.07}$     \\
Line (keV)						 & ...			   & ... 		      & ...		         & 6.67$^{+0.06}_{-0 .05}$	 & 6.67$^{+0.06}_{-0.05}$     \\
$\sigma_{\rm Line}$ (keV)					 & ...			   & ... 		      & ...		         & 0.12$^{+0.09}_{-0.08}$	 & 0.12$^{+0.09}_{-0.08}$     \\
f$_{\rm x}$ (erg\,cm$^{-2}$\,s$^{-1}$)		 & 2.33$\times$10$^{-12}$  & 2.42$\times$10$^{-12}$   & 1.46$\times$10$^{-11}$   & 2.79$\times$10$^{-12}$	 & 2.83$\times$10$^{-12}$     \\
$\chi^{2}_{\nu}$/d.o.f.$^{\mathrm{b}}$		 & 1.01/249		   & 0.94/247 		      & 0.92/246	         & 0.97/246			 & 0.94/244		      \\

\hline                                  \end{tabular} 
\begin{list}{}{}
\item[$^{\mathrm{a}}$] emission measure of the thermal component as numbered; in parenthesis assuming the largest distance (see Section \ref{opticalspectroscopysection});
\item[$^{\mathrm{b}}$] degrees of freedom.

\end{list}
\end{table*}

Spectral fits were carried out simultaneously for the EPIC MOS1, MOS2 and pn cameras. Spectra were accumulated from an event list free of high background flares. In each individual spectrum the data were grouped into energy channels containing at least 25 events.

Both \sao\ and \hd\ exhibit a hard X-ray energy distribution (see Fig.~\ref{xrayspectrum}). A strong Fe\,K$\alpha$  emission line is clearly detected in \hd\ at 6.67\,keV with an equivalent width of $\sim$\,700 eV. A similar feature seems also present in \sao.

\begin{figure*}
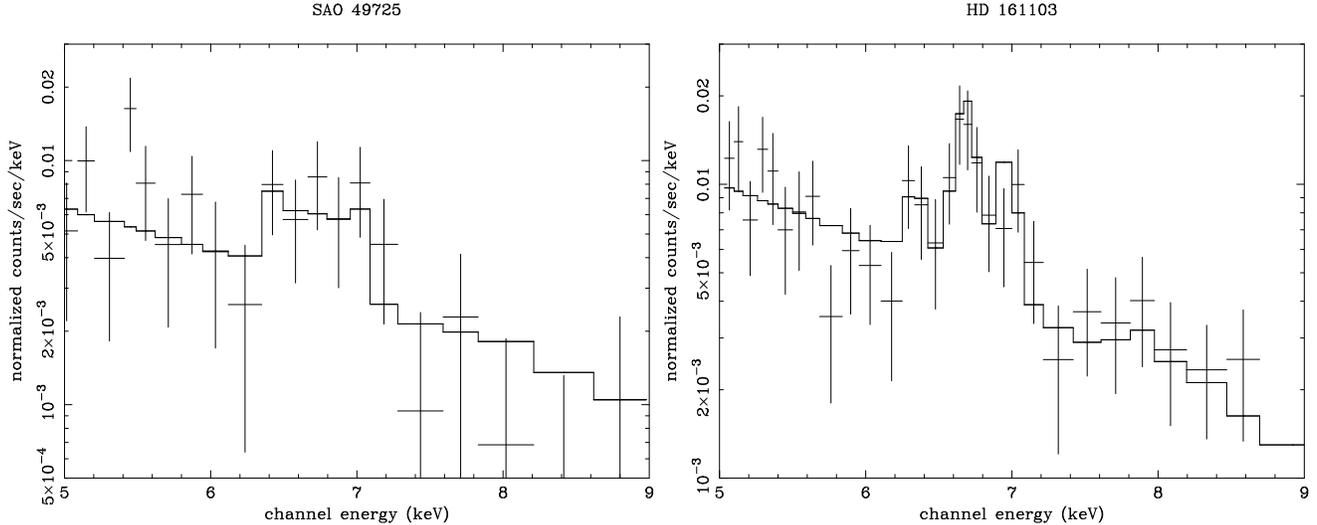
 \centering{ 
\includegraphics[angle=-90,width=8.6cm]{sao49725_allobs_pnpattern0_wabsmekalmekalgauss.ps}
\includegraphics[angle=-90,width=8.6cm]{hd161103_allobs_pnpattern0_wabsmekalmekalgaussian.ps}
\caption{The Fe line complex of \hd, and evidences for its existence in \sao\ seen in the pn single-events spectra accumulating photons during all observation time.
The best fit from {\it wabs}*({\it 2-mekal}\,+\,{\it Gaussian line}) model are shown.} 
\label{xrayspectrumpnpattern0}  } \end{figure*}

Observed spectra can be equally well fitted by an optically thin plasma ({\it mekal}) with T $\sim$\,10$^{8}$ K and solar abundances or by a {\it power law}\,+\,{\it iron line} model with photon index $\Gamma$\,$\sim$\,1.5--1.8 (see Tab.~\ref{fittingparameters}). Including a low temperature component in these models considerably reduces the excess seen in the $\chi^{2}$-residuals around 0.9\,keV and improves the fit quality in the soft part of the spectrum. Its presence is supported by the F-test statistics with a probability of 6$\times$10$^{-2}$ for \sao\ and 6$\times$10$^{-5}$ for \hd\  (one-{\it mekal}), and 0.2 and 5$\times$10$^{-3}$ ({\it power law}\,+\,{\it iron line}) models. 
Freeing metal abundances does not improve the fit and yields metallicities consistent with solar. We do not find evidences for a sub-solar Fe abundance similar to that observed in \gcas\ \citep{Smith04} at the 90\% confidence level (using the {\it vmekal} model). 

We also tried a model in which each {\it mekal} component is affected by a distinct absorption as for \gcas\ \citep{Smith04} without improving the overall fit quality. However this configuration cannot be excluded since the absorption columns are different at the 90\% confidence level. The soft component of \sao\ could be affected by a hydrogen column less dense than that of its hard component (Tab. \ref{fittingparameters}). An opposite behaviour is seen in \hd, in which the hot component is the less absorbed. For \sao\ the resulting plasma temperatures are compatible with those of the model with only one absorption column. 
For \hd\ the hot plasmas have the same temperatures in the two models, but the soft plasma is colder in the two-absorption model, resulting in a high emission measure and a luminosity higher by a factor $\sim$\,10.
We did not find acceptable solutions with a highly absorbed hot component similar to that observed in \gcas\ by \citet{Smith04}.

In all cases (except for the 2-thermal model of \hd\ with individual absorption components) the corresponding luminosities in the 0.2--12\,keV energy range are $\sim$\,(4--12)$\times$10$^{32}$\,erg\,s$^{-1}$ depending on the assumed luminosity class and the X-ray to bolometric luminosity ratio are $\sim$\,4$\times$10$^{-6}$, almost identical for the two sources.

Interestingly, the thermal models naturally represent the H-like and He-like iron lines strength with the correct intensity.  This strongly argues in favour of a thermal interpretation of the X-ray emission of \hd\ and \sao. The corresponding emission measures (EM; Table\,\ref{fittingparameters}) are dominated by the hot component and compare well with those derived for \gcas\ from {\it Rossi}XTE \citep[4.1$\times$10$^{55}$\,cm$^{-3}$;][]{Smith98} and {\it Chandra} \citep[$\sim$\,5$\times$10$^{55}$\,cm$^{-3}$;][]{Smith04}.

\subsubsection{The iron K$\alpha$ complex}
\label{Felinecomplexsection}

In order to improve the energy resolution of the EPIC pn camera we accumulated single-pixel events spectra over the entire observations of \hd\ and  \sao\, i.e., including time intervals of slightly enhanced background. Spectra were grouped in bins containing a minimum of 15 events and we used the Cash-statistic which is more appropriate for small numbers of counts. 

The iron line in \hd\ appears as a Fe\,K$\alpha$ complex of the neutral line at 6.4\,keV, Fe\,{\small XXV} (6.7\,keV) and Fe\,{\small XXVI} (6.97\,keV). Fits to the full observation time data give results consistent with those in Table \ref{fittingparameters} which were restricted to low background intervals. A hot plasma represents reasonably well the emission of the H-like and He-like lines. A separate Gaussian line is needed to account for the Fe\,K fluorescent line (Fig. \ref{xrayspectrumpnpattern0}). We estimated the contribution of each component, by fitting three Gaussian lines on the {\it power law} representing the underlying continuum.
The lower signal to noise ratio available for \sao\ does not allow us to resolve the Fe line complex, but the broad emission feature (EW\,$\sim$\,920\,eV) is compatible with the presence of the three Fe lines detected in \hd.
Freeing all parameters we obtain the centroid of the Gaussian lines and their estimated equivalent width listed in Table \ref{ewironlines}. There is no evidence for non-solar abundances in any of these two sources.

These results strongly suggests that as in \gcas , where high resolution X-ray spectroscopy confirm it, the X-ray emission of \sao\ and \hd\ is essentially due to a hot thin confined plasma, with a fluorescence feature due to reprocessing in cold matter.

\begin{table} \caption{Equivalent width of the emission iron lines in \hd\ and \sao\ estimated of pn single-events spectra. We use an absorbed {\it power law} model to describe the hard continuum and three Gaussian for each Fe line. Errors are given at 1\,$\sigma$.}              
\label{ewironlines}       
\centering          
\begin{tabular}{c c c}  

\hline
\hline                    

Iron line$^{\mathrm{a}}$ (keV) 	& \multicolumn{2}{c}{EW (eV)} \\
				& \sao\ & \hd\ \\
\hline                    

6.4		     & 105 (+145/-84)	&  75 (+49/-52)        	\\
6.7		     & 183 (+120/-100)  & 301 (+90/-65)    \\
6.97		     & 280 (+165/-147)  & 133 (+63/-66)    \\

\hline                    
\end{tabular} 
\begin{list}{}{}
\item[$^{\mathrm{a}}$] frozen parameters.
\end{list}

\end{table}

\begin{figure*} \centering{ 
\includegraphics[bb=1.2cm 5.9cm 20cm 25.1cm,clip=true,width=8.8cm]{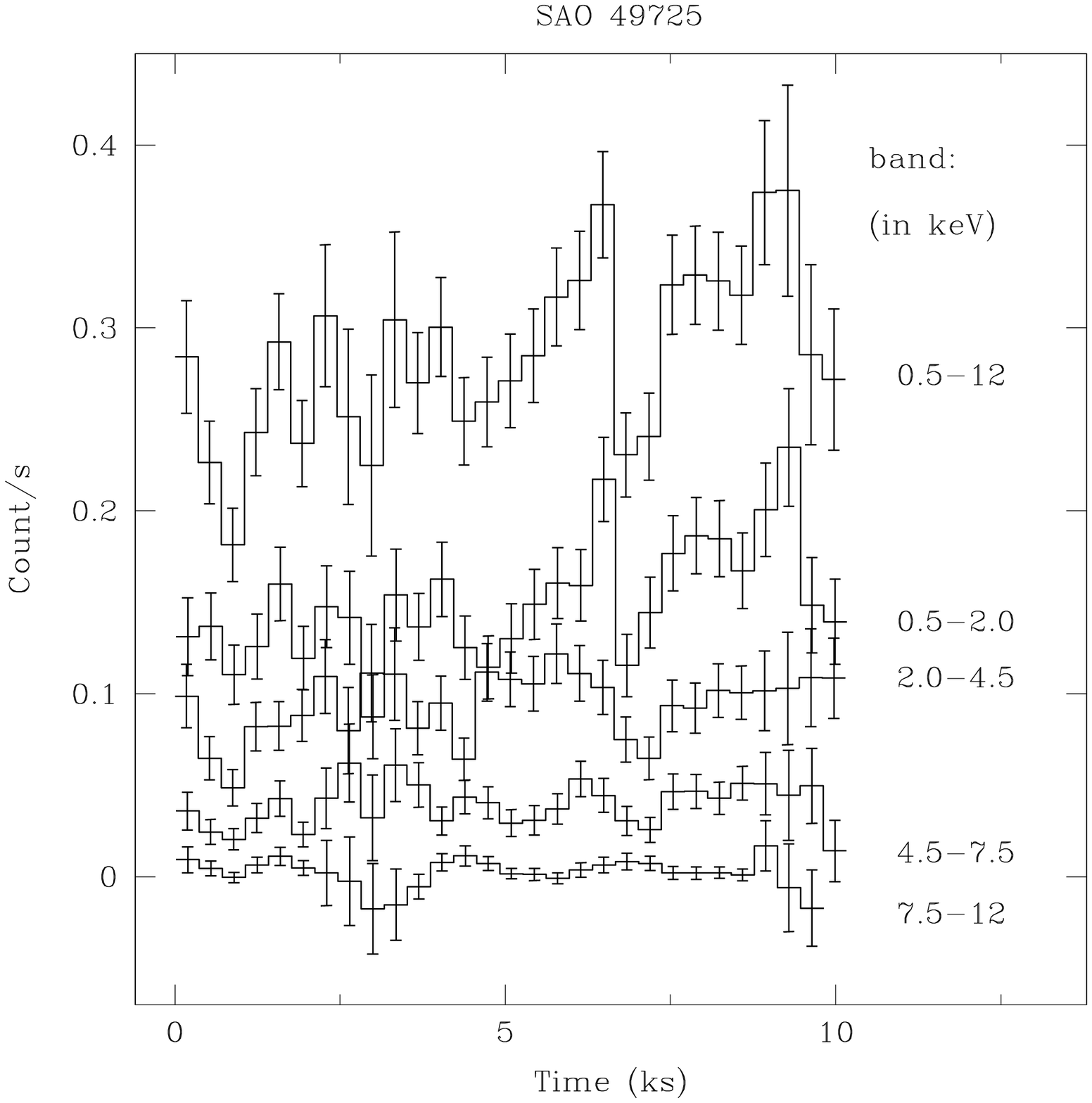} 
\includegraphics[bb=1.2cm 5.9cm 20cm 25.1cm,clip=true,width=8.8cm]{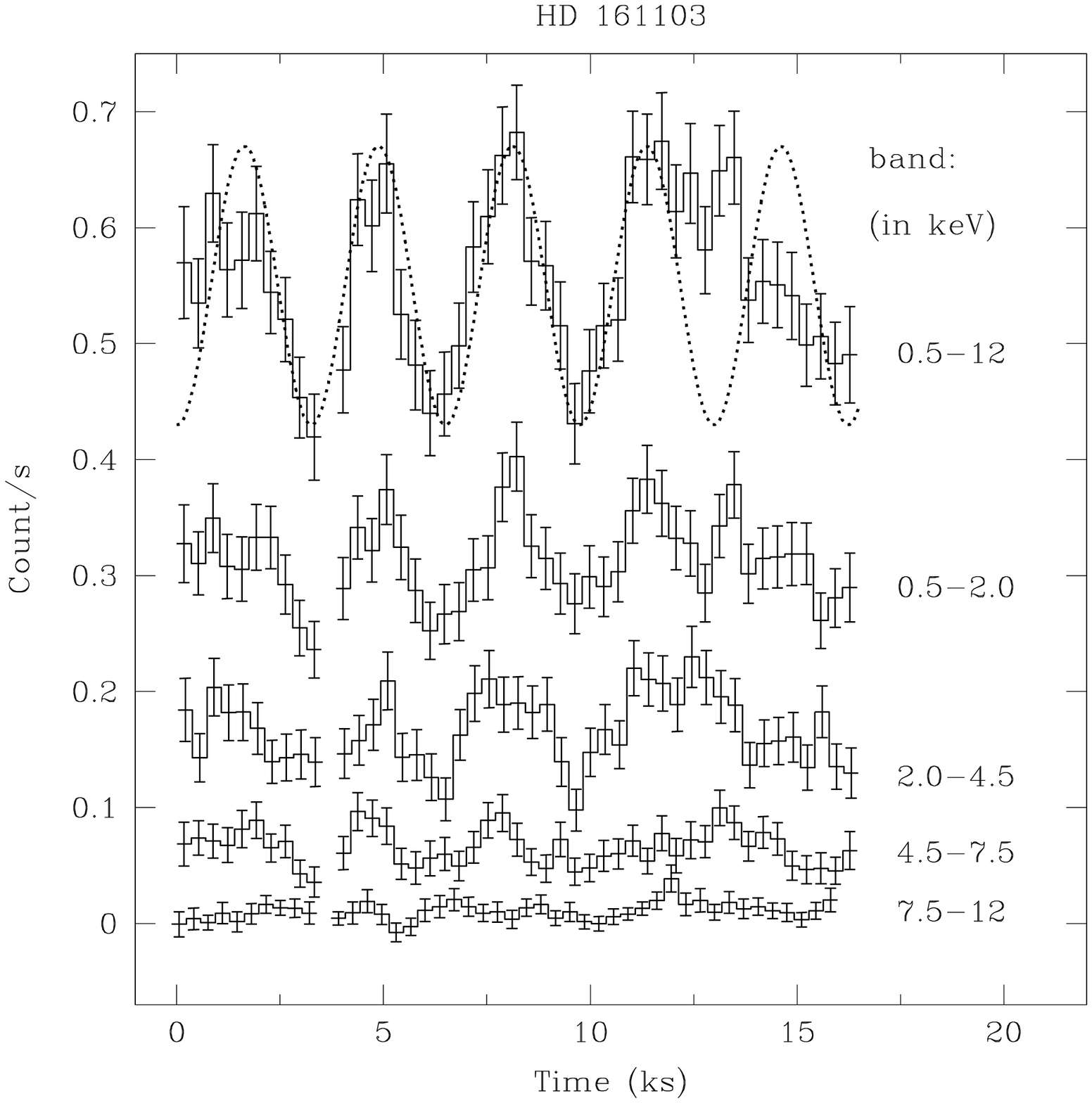}
\caption{EPIC (MOS1+MOS2+pn) background-subtracted barycentric corrected light curves in conventional XMM energy bands. The \hd\ data around t\,$\sim$\,3500 s were lost -- possibly because of a telemetric problem in the satellite. Bin size of 350\,s in all cases.
The dotted curve represents a sinusoidal modulation with period of 3245\,s (see Section \ref{sectionz2n}).
}
\label{lightcurves}} \end{figure*}

\subsection{X-ray light curves}
\label{xraylc}

Both X-ray light curves exhibit significant variability (see Fig.~\ref{lightcurves}). The average X-ray flux of \sao\ slowly increases in the course of the XMM-{\it Newton} observation while showing apparently random variations with a maximum amplitude of $\sim$\,80\% on a time scale of a few hundred seconds. In contrast, the time behaviour of \hd\ seems more regular during the observations. Its intensity varies almost periodically by $\sim$\,60\% on a time scale of $\sim$\,3.2\,ks.

\subsubsection{Spectral variations with intensity}

\begin{figure} \centering \includegraphics[bb=0.5cm 5.5cm 20cm 24.3cm,clip=true,width=8.5cm]{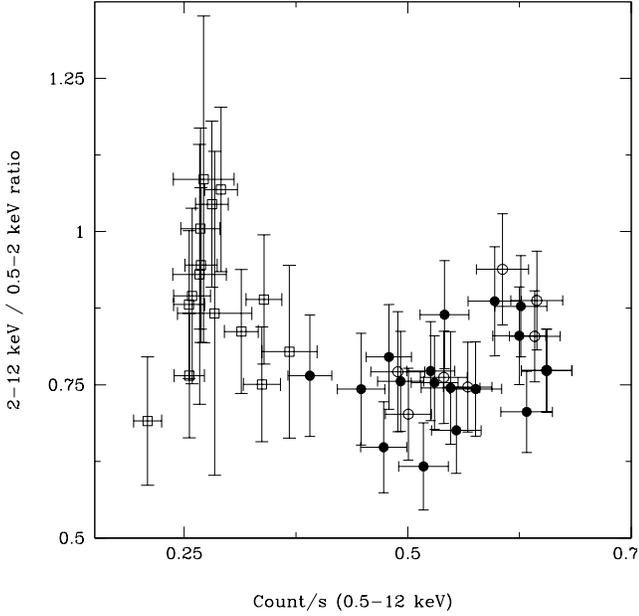} \caption{Hardness-intensity diagram
for \sao\ (open squares) and \hd\ (filled circles for t\,$<$\,12\,ks, open circles for t\,$>$\,12\,ks) using simultaneously all EPIC cameras. The data were grouped in 700\,s time bins.}
\label{hr_intensity_diagram} 
\end{figure}

In order to try to disentangle different possible components contributing to the X-ray emission, we first searched for correlations between the X-ray hardness and intensity of \sao\ and \hd. We adopted the 0.5--2.0 and 2.0--12\,keV energy bands to build a hardness-intensity diagram. Results for the combined EPIC cameras are shown in Fig.~\ref{hr_intensity_diagram}. Although the relatively small hardness ratio changes with X-ray intensity seem significant, the pattern of variability remains unclear. For \hd\ the high state seems to correspond to the hardest energy distribution while an opposite behaviour could be present in \sao. Similar results were obtained using only EPIC pn data.

We further studied spectral changes in \hd, the brightest and the most variable of the two sources, by accumulating
spectra in low and high X-ray flux states for each of the three EPIC cameras. Low and high states correspond to times during which the total 0.5--12\,keV EPIC count rate is below or above of 0.53 count/s respectively (average intensity level; see light curve in Fig.~\ref{lightcurves}). As for other spectra, times of high background were discarded. We also excluded the last part of the observation (t $\ga$ 12\,ks) during which the pronounced 3.2\,ks oscillation seems to vanish. As suggested by the hardness ratio analysis, the X-ray energy distribution of \hd\ appears slightly harder during bright X-ray states, but the best fit parameters in the low and hard flux states are still compatible at the 90\% confidence level (see Tab. \ref{tab_high_low_hd}).

\begin{table} 
\caption{\hd: best fit parameters for the high and low state X-ray spectra. Times of high background  were excluded. We also discard the last missed oscillation (t $>$ 12\,ks in Fig. \ref{lightcurves}). All thermal models assume solar abundances. Quoted errors are at the 90\% confidence level. Fluxes are given unabsorbed in the 0.2--12\,keV energy band. 
}              

\label{tab_high_low_hd}       
\centering          
\begin{tabular}{c c c}            
\hline                    
\hline
& Low & High \\
\hline                    

{\it Mekal}~~~~~~~~~~~~~~~~~~~~~~~~~~~~~~~ &&\\
&&\\
N$_\mathrm{H}$ ($\times$10$^{22}$\,\mbox{cm$^{-2}$})		& 0.26 $^{+0.05}_{-0.05}$ 	 	& 0.30 $^{+0.04}_{-0.03}$ 	\\
$k$T  (\mbox{keV})						& 7.38 $^{+3.55}_{-1.47}$	 	& 8.28 $^{+2.62}_{-1.78}$ 	\\
f$_{\rm x}$ (10$^{-12}$\,\mbox{erg\,cm$^{-2}$\,s$^{-1}$})	& $\sim$\,1.8				& $\sim$\,2.6			\\
$\chi^{2}_{\nu}$/d.o.f.						& 1.06/66			 	& 1.24/110		  	\\

\hline                    

{\it Power law\,+\,Gaussian line} &&\\
&&\\
N$_\mathrm{H}$ ($\times$10$^{22}$\,\mbox{cm$^{-2}$})		& 0.33 $^{+0.08}_{-0.07}$		& 0.34 $^{+0.05}_{-0.04}$	\\
Photon index  							& 1.74 $^{+0.19}_{-0.18}$		& 1.70 $^{+0.12}_{-0.11}$	\\
Line (keV)							& 6.67$^{\mathrm{a}}$			& 6.67$^{\mathrm{a}}$		\\
$\sigma_{\rm Line}$ (keV)					& 0.12$^{\mathrm{a}}$			& 0.12$^{\mathrm{a}}$		\\
f$_{\rm x}$ (10$^{-12}$\,\mbox{erg\,cm$^{-2}$\,s$^{-1}$})	& $\sim$\,2.0				& $\sim$\,3.0			\\
$\chi^{2}_{\nu}$/d.o.f.  					& 1.07/65				& 1.18/109			\\
						
\hline                                  \end{tabular} 
\begin{list}{}{}
\item[$^{\mathrm{a}}$]frozen as integrated spectrum (Tab. \ref{fittingparameters}).
\end{list}
\end{table}
\subsubsection{A search for pulsations}
\label{sectionz2n}

Timing analysis was performed using the Z$^{2}_{\rm n}$ Rayleigh periodogram \citep{Buccheri83} on EPIC background corrected light curves (1\,s, 10\,s and 350\,s time bins) and directly on the EPIC pn event lists. In all cases we corrected times to the barycentric system and accumulated counts in the 0.5--12\,keV energy range. The upper frequency limit is fixed by the acquisition time of 48 ms in the {\it large window} mode used for the pn camera (f $\la$ 10.4 Hz). We applied the Z$^{2}_{\rm n}$ test in $\delta$f steps oversampling the observation window by a factor 1000 ($\delta$f = 0.001/T, where T is the total analyzed observation time).  We fail to detect high frequency pulsations in any of the two sources, but the long period oscillation clearly visible in the light curve of \hd\ is well detected at a period of \,3245\,($\pm$350)\,s  with a pulse fraction of 24\% (see Figure \ref{lightcurves}). The upper limit on the pulse fraction of any other periodic modulation is $\sim$\,10\% \ for \hd\ and \sao.

\section{Discussion} \label{discussion}

The X-ray and optical properties of \sao\ and \hd\ are very similar and can be altogether summarized as follows: 

(i) very hard X-ray spectrum, equally well fitted by a hot plasma ($k$T\,$\sim$\,7--12\,keV) or by a {\it power law}\,+\,{\it iron lines} ($\Gamma$\,$\sim$\,1.5--1.8) model, with in both cases evidences for the presence of an additional soft thermal component ($k$T\,$\sim$\,0.8\,keV); 

(ii) presence of Fe\,K, He-like and H-like iron lines (clearly detected in \hd, and probable in \sao); 

(iii) X-ray (0.2--12\,keV) luminosity of $\sim$\,(4--12)$\times$10$^{32}$\,erg\,s$^{-1}$ implying an X-ray to bolometric luminosity ratio of $\sim$
4$\times$10$^{-6}$;

(iv) variable emission on relatively short time scales (possible stable period of $\sim$\,3200\,s for \hd); no evidence for high frequency pulsations;

(v) B0.5\,III-Ve spectral type;

(vi) intense and quasi-symmetrical Balmer emission profile, suggesting dense and/or very large and apparently stable circumstellar envelopes.

The common properties of these two objects strikingly resemble those of \gcas, an enigmatic early type B0.5\,IVe star first spectroscopically observed in 1866 \citep{Secchi67}.  Recent {\it Chandra} observations of \gcas\ \citep{Smith04} show that the emission spectrum results from different plasmas: a hot-thermal dominant component ($k$T\,$\sim$\,11--12\,keV) with low Fe abundances ($\sim$\,0.22 $\times$ solar), and two or three colder and less absorbed components ($k$T\,$\sim$\,0.1--3\,keV) revealed by the presence of lower excitation lines from Fe L-shell and O\,{\small VII-VIII}. The Fe\,K$\alpha$ complex is present.
The X-ray luminosity (2--10\,keV) of \gcas\ is $\sim$\,5$\times$10$^{32}$\,erg\,s$^{-1}$ \citep{Kubo98}. No pulsation is detected, but there are X-ray flux variations on time scales ranging from seconds to hours \citep{Parmar93,Haberl95,Smith98,Owens99}. In spite of several multi-wavelength observing campaigns, the true nature of the source remains elusive. 

Based on the similarities of \gcas\ with Be/X-ray transient pulsars in the
quiescent state, the first scenario proposed was that of a compact object accreting matter from the circumstellar disc of the Be star \citep{Marlborough78,Murakami86,Haberl95,Kubo98}. The absence of pulsations has cast some doubts on this model, although a long rotation period, a faint magnetic field or geometrical effects could explain the lack of detection. 
Radial velocities variations revealed the presence of an unseen companion star in a $\sim$\,205 day orbit with a mass close to 1\,\Msol\ \citep{Harmanec00,Miroshnichenko02} but it is yet unclear whether the X-ray features observed in \gcas\ could be consistent with an accretion model in such orbit \citep[e.g.][]{Robinson00}.

Assuming that accretion on a companion star is not at work in \gcas, the X-ray emission could be produced by magnetic field recombination in the circumstellar disc or by fields interconnecting the star and the disc \citep{Smith99}. Dynamo effects in the upper stellar atmosphere or circumstellar disc could account for the magnetic field \citep{Robinson02}.  

We discuss below the observed properties of \sao\ and \hd\ in the light of single star and accreting binary
hypothesis.

\subsection{Single stars?}

\begin{figure}
\centering \includegraphics[bb=1.9cm 6.4cm 19.9cm 24.8cm,clip=true,width=8.5cm]{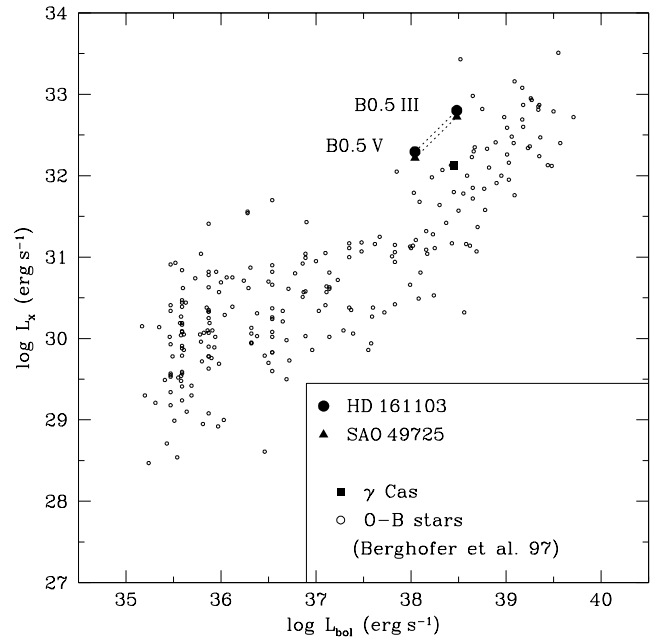} \caption{L$_{\rm x,0.1-2.4\,keV}$\,-\,L$_{\rm bol}$ diagram. \sao\ and
\hd\ are compared with O-B stars at ROSAT energy range. Because of their very hard spectra the X-ray luminosity excess of the new \gcas-like objects would be significantly larger in the 0.2--12\,keV energy band.}
\label{lxlboldiagram} 
\end{figure}

The relatively modest soft X-ray luminosities radiated by \sao\ and \hd\  places these objects at the boundary between early type stars and Be/X-ray transients in quiescent states as well as persistent low-luminosity Be/X-ray systems (X\,Per-like). Their 0.1--2.4\,keV X-ray to bolometric luminosity ratio is clearly above those of the vast majority of normal O-B stars as given in \citet{Berghofer97} (see Fig.~\ref{lxlboldiagram}). However, their much harder X-ray spectra would very significantly move them up in the \Lx /\Lbol \ diagram if the 0.2--12\,keV band were used instead of the softer ROSAT energy range.
Normal O-B stars are soft X-ray emitters with a low temperature plasma of $\sim$\,0.5\,keV \citep{Berghofer96} and moreover usually display no variability on minute time scales and little on longer intervals \citep{Meurs92,Cohen97,Cohen00}. The spectral hardness, luminosity and variability of the X-ray emission of \sao\ and \hd\ are clearly incompatible with wind shock emission of a normal O-B star.

The most relevant single star scenario applicable to \sao\ and \hd\ is probably that proposed by
\citet{Smith04} for \gcas . Based on high resolution X-ray spectroscopy from the {\it Chandra} observatory, these authors suggested a picture in which X-ray flares occur close to the surface of the Be star in magnetically confined plasmas and are partially absorbed and reprocessed in colder material located in the disc. This mechanism is supported by a pattern of correlations between optical, UV and X-ray continuum variations and line features such as Fe and Si\,K fluorescence lines \citep{Smith98,Cranmer00,Robinson02,Smith06b}. 
In \gcas, and also in HD\,110432 (analog system; see Section \ref{sectionnewclass}), migrating subfeatures are seen in the line profile of optical lines supporting the existence of corotating clouds located close to stellar surface and likely magnetically confined \citep[see][]{Smith06a}, and also in agreement with the \citet{Smith04} model.
According to this model, the energy distribution seen in \hd\ and \sao\  could be explained by the sum of a hard dominant and a soft absorbed thin thermal component, the latter representing the normal B star shocked wind emission. 

The photoelectric absorption seen in the X-ray spectrum of \sao\ and \hd\ (N$_\mathrm{H}$\,$\sim$\,3--4$\times$10$^{21}$\,cm$^{-2}$) is fully consistent
with that derived from the interstellar reddening of the Be stars (E(B--V)\,$\sim$\,0.6--0.7) estimated from optical photometry. 
This implies that no strong local absorption occurs.
 
Herbig Ae/Be -- intermediate-mass pre-main-sequence stars, were detected with X-ray luminosities of log\,L$_{\rm x}$\,(erg s$^{-1}$)\,$\sim$\,30--32 and corresponding log\,L$_{\rm x}$/L$_{\rm bol}$ from -7 to -4 \citep{Skinner04,Hamaguchi05}. Their X-ray spectra are usually harder than those of O-B stars and may reach thin thermal temperatures of up to 5\,keV, approaching those of \sao\ and \hd\ but remaining significantly cooler. They often exhibit a Fe\,K$\alpha$ line. A last similarity with the two targets studied in this paper is their significant time variability, sometimes in the form of flare-like events. There is no evidence, however, that the targets are located near star forming regions.

\subsection{Accreting binaries?}

The relatively small (typically a factor 10 in the 0.1-2.4\,keV band) excess of hard X-ray luminosity above the usual emission of normal O-B stars can be naturally explained by accretion onto a compact object. Several distinct mechanisms can account for the low observed X-ray luminosity which is in general lower than those observed from classical Be/X-ray transients in quiescence and persistent low-luminosity Be/X-ray binaries. First, the mass accretion rate itself can be smaller than in other massive X-ray binaries. This can happen if the local density of matter is small, i.e., if the neutron star is located outside of the circumstellar disc on a tilted orbit and is only fed by the high velocity wind, or if disc truncation mechanisms are extremely efficient \citep{Okazaki01}. Alternatively, the mass accretion rate may be similar to those encountered in Be/X-ray systems, but the accretion potential well is weaker. This will be the case if the magnetized neutron star is in the propeller regime \citep{Stella86}, where accreted matter is stopped at the magnetosphere or if the compact object is a white dwarf. 

Although the power-law indices of the energy distribution of \sao\ and \hd\ compare well with those of classical Be+NS transients, in such systems the iron emission, when present, is dominated by the fluorescence line at 6.4\,keV with weak or no He-like and H-like component.  The probable thermal nature of the X-ray emission of \sao\ and \hd\ is also in disagreement with the non-thermal energy distributions which are always detected in accreting neutron stars.  
Interestingly, several cataclysmic variables with comparable X-ray luminosities display similar X-ray spectra with high $k$T and the same iron complex in emission \citep[see e.g][]{Hellier04,Pandel05}. This similarity has been considered as one of the main arguments for the presence of an accreting white dwarf in \gcas\ \citep{Haberl95,Kubo98}.

Apart from the fact that the X-ray luminosity seen in \gcas\ is apparently incompatible with that expected from an accreting WD in a 205 day orbit \citep{Robinson00}, one of the often invoked objections to the Be+WD hypothesis for \gcas\ is the difficulty to avoid making a neutron star in the course of a binary evolution ending with a B0.5\,IV star. However, the long lasting mass transfer phases occurring during the evolution of close binaries strongly influences the evolutionary tracks and the final stellar products. In fact the strong dependence of the history of the system on initial conditions leads to a very large range of configurations. \citet{Pols91} show that Be+WD have a distribution peaking around the B2 spectral type, but are still possible for stars earlier than B0.5. In $\phi$\,Per, there is a B0.5\,Ve star (9.3\,\Msol) orbited by a hot sub-dwarf with $M$\,=\,1.1\,\Msol \citep{Gies98}. 
In the absence of further mass transfer, such sub-dwarf will become a white dwarf and not a neutron star. There are also at least three known B+WD binaries: HR\,2875 \citep[B5\,V;][ - triple system]{Motch97,Vennes97,Burleigh98,Vennes00}, $\theta$\,Hya \citep[B9.5\,V;][]{Burleigh99} and 16\,Dra \citep[B9.5\,V;][]{Burleigh00}.
Three possible endpoints of WD\,+\,massive star systems are the binary pulsars PSR\,B2303+46 \citep{vanKerkwijk99}, PSR\,J1141-6545 \citep{Kaspi00} and PSR\,J1906+0746 \citep{Lorimer05}.

\subsubsection{Accretion regimes}
\label{Accretionregimes}

The accretion regime on a neutron star, either onto its surface or its magnetosphere, depends on the relative values of the magnetospheric ($r_{m}\propto B_{0}^{4/7}R_{NS}^{12/7}M_{NS}^{-1/7}\dot{M}_{acc}^{-2/7}$) and corotation ($r_{c}\propto M_{NS}^{1/3}P_{s}^{2/3}$) radius \citep{Stella86}. Accretion on the magnetosphere (\rmag\,$>$\,\rc) was proposed to explain the quiescent states (\Lx\,$\sim$\,10$^{32-36}$\,erg\,s$^{-1}$) in high mass X-ray binaries such as A0538--66, 4U\,0115+63 and V0332+53 \citep{Corbet97,Campana01,Campana02}. \citet{Corbet97} detected an emission iron line in A0538-66 from ASCA observations during a quiescent state ($\sim$\,5.5$\times$10$^{36}$\,erg\,s$^{-1}$), which was not present during high emission states. This feature was interpreted as a single line at 6.54\,keV, or as a blend of 6.4 and 6.7\,keV lines, and was considered as a possible signature of magnetospheric accretion.

\citet{Corbet84} noted that Be/X-ray binaries are located in a relatively narrow locus in the \Ps\ {\it versus} \Porb\ diagram which could be explained as the region of spin equilibrium where \rmag\,$\simeq$\,\rc\ \citep{Waters89b}. In this scenario the accretion regime continuously switches between accretion onto the surface, accretion on the magnetosphere, as well as non accreting states.

The accretion luminosity can be estimated by $L_{acc}(r)=\eta GM\dot{M}_{acc}/$$r$, where \Macc\ is the accretion rate onto an object of mass $M$, in which potential-radiative energy conversion occurs with an efficiency $\eta$ at a distance $r$ of the centre of the compact object.  When accretion occurs on the neutron star (white dwarf) surface, $r$ = $R_{NS}$ (or $r$\,=\,$R_{WD}$). If accretion occurs on the magnetosphere $r$\,=\,\rmag, and the resulting luminosity can be estimated by:

\begin{displaymath}
{L_{acc}^{mag}\simeq 1.1\times10^{33}(\frac{M_{NS}}{1.4\,M_{\odot}})^{8/7}[(\frac{B_{NS,0}}{3\times10^{12}\,G})(\frac{R_{NS}}{10\,km})^{3}]^{-4/7}}
\end{displaymath}

\begin{equation}
\label{magnethospericluminosity}
\,\,\,\,\,\,\,\,\,\,\,\,\,\,\,\,\,\,\,\,\,\,\,\,\,\,\,\,\,\,\,\,\,\,\,\,{\,\,\times(\frac{\dot{M}_{acc}}{10^{-10}\,M_{\odot}\,yr^{-1}})^{9/7}\,\,\,{\rm (in\,erg\,s^{-1}).}}
\end{equation}

The transition between accretion on the magnetosphere and accretion on the neutron star ideally occurs at an accretion rate (\Macclim) for which \rmaglim\,=\,\rc\ and should be accompanied by a several orders of magnitude jump in X-ray luminosity \citep{Corbet96}. 

\begin{figure} \centering \includegraphics[bb=1cm 5.2cm 19.5cm 25.5cm,clip=true,width=8.5cm]{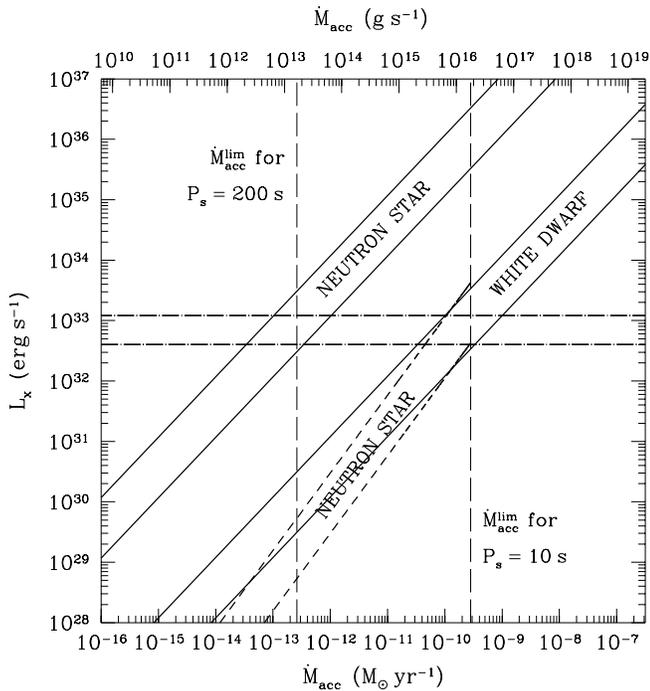} \caption{Luminosity -- accretion rate diagram for accretion onto the white dwarf and neutron star surfaces (solid lines, as indicated), and for magnetospheric accretion onto a neutron star (dashed lines) -- upper and lower lines refer to $\eta$\,=\,1 and 0.1, respectively. Horizontal lines show the observed X-ray luminosity (0.2--12\,keV) of \hd\ and \sao. Vertical lines mark \Macclim\ for a neutron star with \Ps\,=\,10 and 200\,s (see text).}
\label{luminosity_accretionrate} 
\end{figure}

Figure \ref{luminosity_accretionrate} shows $L_{acc}$ {\it versus} \Macc\ for three cases:  accretion onto the neutron star surface, accretion onto the neutron star magnetosphere and accretion onto a weakly magnetized white dwarf. We adopt a bipolar description for the magnetic field of the neutron star with $B_{NS,0}$\,=\,3$\times$10$^{12}$\,G at the pole; masses and radii of 1.4\,\Msol\ and 10\,km for the neutron star and 1.0\,\Msol\ and 0.01\,\Rsol\ for the white dwarf. We assume accretion efficiencies $\eta$ of 1.0 and 0.1.

The observed X-ray luminosities of \hd\ and \sao\ ($\sim$\,(4--12)$\times$10$^{32}$\,erg\,s$^{-1}$ at 0.2--12\,keV) could thus be produced by direct accretion on a neutron star surface for an accretion rate of $\sim$\,2$\times$10$^{-13}$\,\Msol\,yr$^{-1}$ corresponding to \rmag\,$\sim$\,6$\times$10$^{9}$\,cm. This requires the opening of the magnetic barrier, i.e., neutron star rotation periods larger than $\sim$\,200\,s. The same X-ray luminosity could be explained by accretion on the magnetosphere of a neutron star with a spin period shorter than $\sim$\,10\,s corresponding to \rmag\,$\sim$\,8$\times$10$^{8}$\,cm or onto a white dwarf assuming a much higher accretion rate of $\sim$\,2$\times$10$^{-10}$\,\Msol\,yr$^{-1}$.

\subsubsection{The likelihood of the accretion scenario}

Estimating X-ray luminosities consistent with the various accretion regimes requires a good understanding of the density and velocity fields of the circumstellar matter along the orbit of the accreting star. 
 
A growing number of observational evidences and theoretical developments have reinforced the idea that the viscous decretion disc model proposed by \citet{Lee91} can explain most of the disc-related observational properties of Be stars. The major step forward was the recognition that radial outflow velocities were highly subsonic in the whole line emitting disc region. Tangential velocities are mainly Keplerian and the viscosity acts in an opposite way as in an accretion disc, transporting the angular momentum of the inner regions toward the outer disc. This scenario has been successful in explaining the observed Balmer line profiles \citep{Hanuschik00} and the IR colours \citep{Porter99}. \citet{Negueruela01} have shown that the tidal torque of a stellar companion can efficiently truncate the viscous disc at a resonance radius which depends on disc viscosity, on the mass ratio of the stars and on the orbital parameters. \citet{Okazaki01} have applied this model to the class of Be/X-ray transients and have shown that many of the X-ray and optical behaviour of these transients (e.g. V/R variations, occurrence of type I or II outbursts) could be explained in this framework. Disc truncation  should be more efficient in systems with small orbital eccentricities and small orbital periods \citep{Zhang04} and could deplete the circumstellar disc at relatively small distances from the Be star. 

Unfortunately, the decretion disc model makes no prediction on the density and velocity fields beyond the truncation radius. Even in the absence of truncation, the density and velocity laws at the outer edge of the disc, beyond the trans-sonic point are ill defined. At a large distance of the star, it is reasonable to think that the pressure of the high velocity stellar wind will considerably change the flaring disc structure \citep{Kubat06} dragging to higher velocities the disc material. 

Direct accretion on a neutron star implies rather extreme conditions. First, the spin period must have slowed down to large values albeit not unusual, in order for accretion to take place. Second, the orbital period must be long enough to explain the small accretion rate. Low luminosity persistent Be/X-ray sources such as X Per (\Porb\,=\,250$^{d}$) or RX\,J0146.9+6121 with X-ray luminosities of the order of 10 to 100 times that observed from \sao\ or \hd\ usually do not display any iron line \citep{Mereghetti00,Coburn01}. The absence of fluorescence iron line probably reflects the scarcity of the circumstellar medium at the large distances where the accreting neutron star orbits. In this respect, the strength of the iron line in \hd\ and \sao\ rather suggests that the X-ray source is embedded in a relatively dense cold medium, casting severe doubts on the accreting neutron star hypothesis.

Accretion onto a white dwarf (or onto the magnetosphere of a fast spinning neutron star) requires much higher accretion rates. If the X-ray luminosity scales with the gravitational potential well we expect the hypothetical white dwarf to be on the same kind of orbit than neutron stars with X-ray luminosities of the order of 10$^{35-36}$\,erg\,s$^{-1}$, comparable to those of some quiescent states of A\,0535+26 for instance \citep{Motch91}. A\,0535+26 displays a rich behaviour with regular type I outbursts occurring close to periastron passage, rare type II outbursts and undergoes long periods of low X-ray luminosity during which no outburst occurs at periastron. Disc truncation can adequately explain these contrasted variability regimes \citep{Okazaki01}. Because of the absence of birth kick velocity, white dwarf eccentricities should be significantly smaller than those of neutron stars and as a consequence, disc truncation efficiency should be much higher, leading to smaller discs. Replacing the neutron star in A\,0535+26 in a non-outbursting state by a white dwarf could perhaps explain the mean X-ray luminosity and the apparent absence of type I or II outbursts in \gcas\ like objects. In general, an orbiting white dwarf may truncate the Be disc so efficiently that it will only accrete from matter escaping beyond the truncation radius in agreement with the low X-ray luminosities and the apparent absence of outbursts. However, this scenario has to face a major difficulty, namely that optical observations show that the circumstellar discs in \gcas\ and also probably in its twins are large and stable in contradiction with what is expected from truncated discs. 

\subsubsection{Be\,+\,black hole systems?}

Black hole transients in quiescent states (BHQ) have power law spectra with photon index ranging between 0.9 and 2.3 (mean index of $\sim$\,1.5) and luminosity (0.3--7\,keV) of $\sim$\,10$^{30-33}$\,erg\,s$^{-1}$ \citep{Garcia01,Kong02,Hameury03}. These power law index are in principle similar to those derived for \sao\ and \hd, albeit Fe\,K$\alpha$ complex is not seen in BHQ. The black hole hypothesis is ruled out by the orbital solution proposed for \gcas\ \citep{Harmanec00} but may still work for our targets. As suggested by \citet{Zhang04} the effective truncation of the decretion disc around the Be star implies long quiescent phases of low X-ray luminosity, if the accreting black hole is in a narrow orbit ($\la$\,30 days) of low eccentricity.
But the stability of the circumstellar disc might not be consistent with this scenario.

\subsection{What is the nature of the 3200\,s oscillation in \hd?}

The short duration of the X-ray observation does not allow to constrain the stability of the 3200\,s period detected in the X-ray light curve of \hd. Such a long period can be due to a spinning white dwarf or neutron star and its confirmation would be a major step forward in the understanding of these systems. In the case of 2S\,0114+650 a $\sim$\,2.78-h oscillation is detected which probably reflects the spin period of a neutron star emitting an X-ray luminosity of \Lx\,$\sim$\,10$^{35}$\,erg\,s$^{-1}$ \citep[2--10\,keV;][]{Finley92}.

A similar quasi-periodic time behaviour is often observed in \gcas\ \citep{Parmar93,Haberl95,Smith98,Owens99} with comparable time scales. This variability has been interpreted in terms of flare-like events similar to those detected from the hot coronae of solar-like stars \citep{Smith04}. This quasi periodic flaring behaviour also compares well with that of the Be/X-ray binary EXO\,2030+375 \citep{Parmar89}, in which a series of six X-ray flares with duration between 80 and 130\,min was observed with recurrence times of $\sim$\,230\,min. A last possible analogy is the flaring activity on timescales of $\sim$\,1\,h detected in 4U\,2206+54 by \citet{Masetti04}. 

Although $\beta$ Cephei stars exhibit optical photometric pulsations with comparable periods of the order of 3--6 hours, they usually display little X-ray variability \citep{Cohen00,Cassinelli94} and are thus unlikely to be related to the phenomenon we observed.

\section{A new class of X-ray emitters}
\label{sectionnewclass}

Currently four Be stars with very similar X-ray and optical properties to \gcas, \hd\ and \sao\ have been serendipitously discovered in the XMM-{\it Newton} surveys \citep[see][for a recent review]{Motch06a}. 
Examples are USNO\,0750--13549725 and SS\,397 \citep{Motch03,Motch06a,Motch06b}. The first object is a blue straggler B1\,IIIe star at the centre of NGC\,6649 \citep{Marco06} -- an open cluster distant of $\sim$\,1.6\,kpc and heavily reddened (A$_{V}$\,$\sim$\,4.95), with an estimated age of 50 Myr \citep{Turner81,Walker87}. Its X-ray spectrum can be fitted by a {\it power law} model with $\Gamma$\,$\sim$\,1.4--1.6 or by a {\it mekal} model with $k$T\,$\sim$\,10\,keV. The resulting X-ray (0.2--12\,keV) luminosity is $\sim$\,3$\times$10$^{32}$\,erg\,s$^{-1}$. The second object, SS\,397, is a B0.5\,Ve star and also a hard X-ray emitter. The temperature as inferred by a {\it mekal} model fit is $k$T\,$\sim$\,13\,keV, or $\Gamma$\,$\sim$\,1.6--1.8 from {\it power law} model. An emission iron line, or blend of the Fe\,K$\alpha$ complex, is also observed. Strong flux variations in both  USNO\,0750--13549725 and SS\,397 were detected by XMM-{\it Newton} observations, on very similar time scales as here found in \hd\ and \sao\ (Section \ref{xraylc}). 
Detailed results on these new sources will be presented elsewhere. A {\it Beppo}SAX observation of the HEAO-1 Be/X-ray binary candidate HD\,110432 (B0.5\,IIIe) was reported by \citet{Torrejon01}. Its X-ray spectrum was also well described by a thermal hot plasma of $k$T\,$\sim$\,11\,keV -- resulting in an unabsorbed luminosity (2--10\,keV) of $\sim$\,3.4--7$\times$10$^{32}$\,erg\,s$^{-1}$. A possible periodicity of $\sim$\,14\,ks was suspected.
As in \gcas, optical spectroscopy of HD\,110432 revealed migrating subfeatures in the line profiles suggesting the existence of magnetically confined circumstellar clouds corotating close to the star's surface \citep[see results and discussion in][]{Smith06a}.  

It is now clear that the long standing \gcas\ puzzle is not an unique case anymore. Several new 
\gcas\ analogs are found in X-ray surveys. 
Amazingly their X-ray and optical properties fall in a very narrow range, in particular their spectral types which are all consistent with $\sim$\,B0.5e.    
They constitute {\it a new and well defined class of X-ray emitters} that is characterized by: [1] a hard X-ray emission likely thermal and dominated by a component with 7\,$\la$\,$k$T\,(keV)\,$\la$\,13 as supported by the presence of a prominent Fe\,K$\alpha$ complex, [2] with a variable behaviour, and [3] moderate 0.2--12\,keV luminosity (32\,$\la$\,log\,L$_{\rm x}$\,(\ergs)\,$\la$\,33); and from their optical properties, [4] a large and probably stable circumstellar disc.

\section{Conclusions} 

Our XMM-{\it Newton} observations of two low luminosity hard X-ray emitting Be stars selected from the ROSAT all-sky survey reveal a very likely thermal emission of high temperature ($k$T\,$\ga$\,8\,keV). 
A strong iron line including a fluorescence component is detected. 
They also share very similar optical properties such as spectral types (B0.5\,III-Ve) and strength of the Balmer emission (EW\,$\sim$\,-31\,\AA).
This pattern of common X-ray and optical properties appears quite similar to that of the so far unique star \gcas\ and points at the emergence of a new class of \gcas\ analogs. 

\hd\ displays a 3200\,s oscillation which could be the signature of a rotating accreting compact object or with equal likelihood the analog of the long period quasi-oscillations seen in the X-ray observations of \gcas. We discuss the possible X-ray emission mechanisms in the framework of the models proposed for \gcas . A single star scenario in which X-ray flares occur close to the stellar photosphere in magnetically confined regions could in principle account for the observations but may be challenged by the low intrinsic absorption of the X-ray source. This model cannot be really proved or falsified with the presently available data. The accretion scenario onto a neutron star or a white dwarf faces a number of difficulties, one of the most severe being the absence of evidence for disc truncation effects. The strength of the 6.4\,keV line strongly argues against accretion onto a neutron star on a long period orbit.

\begin{acknowledgements}

We would like to thank Dr. Myron A. Smith for useful discussions.
The INT is  operated on the island of La Palma by the Isaac Newton Group 
 in the Spanish Observatorio del Roque de Los Muchachos of the Instituto de 
 Astrof\'{\i}sica de Canarias. 
 The G.D. Cassini telescope 
 is operated at the Loiano Observatory by the Osservatorio Astronomico di 
 Bologna. 
 Based in part on observations made at 
 Observatoire de Haute Provence (CNRS), France. Partially based on data 
collected at the European Southern Observatory, Chile (71.D-0151).
 RLO  acknowledges financial support from Brazilian agencies FAPESP (grant 03/06861-6) and CAPES (grant
BEX0784/04-4), and Observatoire de Strasbourg.
IN is a researcher of the 
 programme {\em Ram\'on y Cajal}, funded by the Spanish Ministerio de 
 Ciencia y Tecnolog\'{\i}a (currently Ministerio de Educaci\'on y 
 Ciencia) and the University of Alicante, with partial 
 support from the Generalitat Valenciana and the European Regional 
 Development Fund (ERDF/FEDER). 
 This research is partially supported by the MCyT (currently MEC) under 
 grant AYA2002-00814. 

\end{acknowledgements}

\end{document}